# Aberrations of the point spread function of a multimode fiber

**Adrien Descloux,[1,2,3] Lyubov V. Amitonova,[1,*] and Pepijn W. H. Pinkse[1]**

[1]*Complex Photonic Systems (COPS), MESA+ Institute for Nanotechnology, University of Twente, PO Box 217, 7500 AE Enschede, The Netherlands*
[2]*Current address: LBEN/LOB, School of Engineering, École Polytechnique Fédérale de Lausanne (EPFL), Station 17, 1015 Lausanne Switzerland*
[3]*adrien.descloux@epfl.ch*
[*]*l.amitonova@utwente.nl*

**Abstract:** We investigate the point spread function of a multimode fiber. The distortion of the focal spot created on the fiber output facet is studied for a variety of the parameters. We develop a theoretical model of wavefront shaping through a multimode fiber and use it to confirm our experimental results and analyze the nature of the focal distortions. We show that aberration-free imaging with a large field of view can be achieved by using an appropriate number of segments on the spatial light modulator during the wavefront-shaping procedure. The results describe aberration limits for imaging with multimode fibers as in, e.g., microendoscopy.

## 1. Introduction

Light microscopy has been the key tool for biological and medical research for more than three centuries. Nowadays optical microscopy is routinely used for imaging and for studying a variety of cellular-level processes in living tissue [1]. One of the main limitation of a traditional microscope technique is the imaging depth [2]. High-resolution imaging is possible only up to one millimeter below the tissue surface, because at greater depths, multiple light scattering inherent to almost any biological tissue deteriorates the image [3]. To overcome this constraint, microscopy techniques based on miniature optical probes have been developed.

The advent of cutting-edge fiber-optic components plus the plethora of light microscope techniques now available offer a powerful platform for microendoscopy, improving imaging resolution, field of view and probe miniaturization. Microendoscopy provides minimally-invasive access to deep tissues in living animals [4,5]. Several strategies of optical probe miniaturization based on recent achievements in micro-optic and fiber optic technologies have emerged, including fiber bundles, gradient-index (GRIN) lenses and multimode (MM) fibers, as main component. Fiber bundles constructed from thousands of individual cores allow deep-tissue imaging by using each core as a single pixel [6]. Optical imaging via fiber bundles allows high-speed fluorescence microscopy in freely moving mice [7], long-term deep-tissue measurements of neural activity [8], *in vivo* monitoring of chemically specific markers [9] and optical control of neural circuits [10]. However, fiber bundles suffer from a low imaging resolution dictated by the diameter of the individual fiber cores and the distance between the fibers. Fiber bundles are not suitable for 3D microscopy since imaging is confined to the surface of the fiber bundle. In contrast, GRIN probes provide microscopic resolution in all three dimensions. Miniature GRIN lenses allow to image cells in deep brain regions without pixilation [11,12]. The main problem of GRIN-lens-based microendoscopy is the relatively small field of view (FOV) of the probe. Typical GRIN lenses with a diameter of 1 mm and numerical aperture NA = 0.4 have a FOV of only about 250 μm [13]. That makes the endoscopic probe itself order of magnitude bigger than what is really needed. In addition, the actual resolution of GRIN probe tends to be worse than the theoretical limit, because the effective NA of GRIN lenses is degraded by a factor of 1.5–2 due to spatial aberrations of the lenses [13].

New horizons in microendoscopy have opened up by the advent of complex wavefront shaping techniques, in which light is focused through a scattering medium using spatial light modulators [14–16]. Standard multimode optical fibers can now be used as a thin imaging probe [17–19]. Multimode fibers potentially offer imaging with a better resolution and a minimal cross section compared with the fiber-bundle and GRIN lens approaches, by utilizing and controlling the high number of transverse fiber modes [20]. The resolution of multimode-fiber imaging methods is typically controlled by the fiber probe's numerical aperture. Recently, it was shown that complex wavefront shaping technique together with a properly designed multimode photonic crystal fiber allows high-resolution imaging with an NA of more than 0.6 [21]. Moreover, enhanced resolution beyond the diffraction limit is possible by using saturated excitation and temporal modulation [22]. Recent works also propose methods for high-speed multimode-fiber imaging [23,24].

Despite the broad interest in multimode-fiber imaging, a systematic study of aberrations inherent to this method of microendoscopy remained out of scope. Here, we investigate focus distortions associated to multimode-fiber imaging. We show that the shape of the focused spot on the output fiber facet strongly depends on the parameters used to control the phases of the different fiber modes. We demonstrate that aberration-free imaging with a large field of view can be achieved by using an appropriate number of segments on the spatial light modulator during the wavefront-shaping procedure.

## 2. Theoretical model

In order to gain insight into the propagation of wavefront-shaped light through multimode fibers, the following mathematical model has been developed. In essence, light propagation through a fiber is supported by a finite set of TE, TM and hybrid modes [25]. Solving the wave equation in a cylindrical geometry with known boundary conditions allows to compute the (complex-valued) electromagnetic field distribution $E_n(x,y)$ and the (real-valued) propagation constant $\beta_n$ of all the $n$ modes. In the following we make the weak guidance approximation (i.e. $n_{core} \approx n_{clad}$), where the modes of the fiber simplify into a degenerated set of Linearly Polarized (LP) modes.

The propagation $P\{E_{in}\}$ of any input field $E_{in}(x,y)$ through a fiber of length $L$, producing the output field $E_{out}(x,y)$, can be expressed as:

$$E_{out}(x,y) = P\{E_{in}(x,y)\}|_{z=L} = \sum_{n=1}^{nModes} \alpha_n E_n(x,y) e^{-i\beta_n L}, \quad (1)$$

where $L$ is the length of the fiber, $E_n(x,y)$ the normalized $n^{th}$-LP mode and $\alpha_n$ is the cross-correlation or overlap between the normalized $n^{th}$-mode and the input field, given by:

$$\alpha_n = \iint_{x,y} E_{in}(x,y) \cdot E_n^*(x,y) dy dx. \quad (2)$$

The excitation field is decomposed into the basis of fiber modes. Each mode propagates at its own speed along the length of the fiber and consequently acquires a different phase, creating a seemingly random interference pattern. To gain control over this modal scrambling, a spatial phase control device is placed in the Fourier plane of the proximal end (pupil plane) of the fiber as shown in Fig. 1. The full propagation can be expressed as:

$$E_{out}(x,y) = P\{\mathfrak{F}\{b(k_x,k_y) \cdot e^{i\varphi(k_x,k_y)}\}\}|_{z=L}, \quad (3)$$

where $\varphi(k_x, k_y)$ is the applied pupil spatial phase distribution, $b(k_x, k_y)$ is a Gaussian-shaped illumination pattern and $\mathfrak{F}\{\}$ the 2D Fourier transform. Note that we label the spatial coordinates in the input pupil with $k_x$ and $k_y$ since the pupil is in the Fourier plane of the input facet.

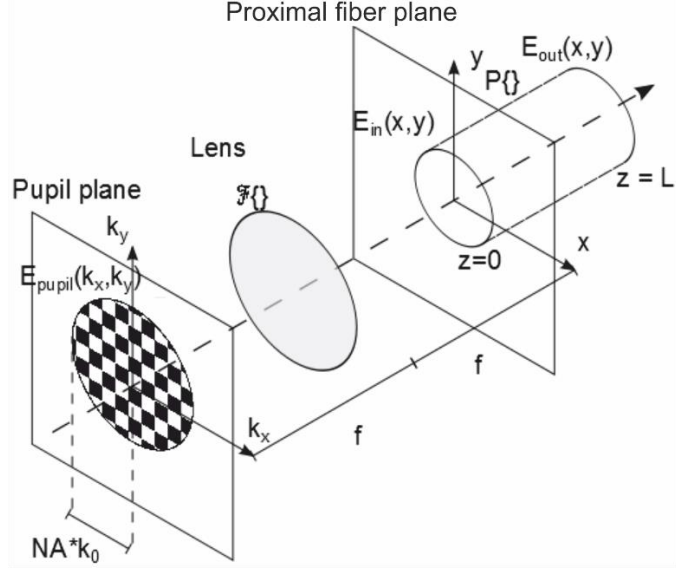

Fig. 1. Scheme of the light propagation in our theoretical model: a wavefront-shaped field $E_{pupil}(x,y)$ is focused onto the fiber core in a 2f configuration. The field $E_{in}(x,y)$ is decomposed into modes and propagates independently through the fiber to form the field $E_{out}(x,y)$.

We consider two different approaches for light focusing through multimode fibers. The first case consists of computing the propagation of the ideal target field from the distal end of the fiber to the pupil plane. Focusing is achieved by back propagation of the resulting field to obtain the corresponding input field. This corresponds to the ideal case, which sets the limits of our system and can be written as

$$E_{\text{pupil ideal}} = \mathfrak{F}^{-1}\left\{P\left\{E_{\text{target}}(x,y)\right\}\big|_{z=-L}\right\}, \tag{4}$$

where $E_{\text{target}}(x,y)$ is the target field distribution. In the simple case where the target field can be expressed as $\delta(x - x_0, y - y_0)$, we obtain $\alpha_n = E^*_n(x_0, y_0)$. The pupil field can then be expressed as:

$$E_{\text{pupil}} = \sum_{n=1}^{nMode} E^*_n(x_0, y_0) e^{i\beta_n L} \mathfrak{F}^{-1}\{E_n(x,y)\}. \tag{5}$$

The pupil field is given by the coherent sum of the phase-shifted Fourier transform of the modes, where modes are living on circles of which the radii are functions of the spatial frequency content of the modes. Therefore, the position of a segment within the pupil space defines the set of modes addressed by that segment. Back propagation of this field is equivalent to a convolution of the target field with the diffraction-limited complex point spread function determined by the wavelength and the numerical aperture of the fiber.

In the second approach, the pupil phase mask is determined by measuring the response of the fiber under specific illumination conditions. In detail, the pupil phase is spatially divided into square segments with a side length equal to $2NAk_0/nSeg$, where $NA$ is the numerical aperture of the fiber, $k_0$ the propagation constant of the light in vacuum and $nSeg$ the number of segments in the $k_x$ or $k_y$ direction. The pupil field can then be represented as a sum of $nSeg^2$ spatially shifted 2D rectangle-shaped window functions with an independent phase:

$$e^{i\varphi(k_x,k_y)} = \sum_{j=1}^{nSeg^2} e^{i\varphi_j} \cdot \text{rect}\left(NAk_0/nSeg, NAk_0/nSeg\right) \otimes \delta(\mathbf{k}-\mathbf{k}_j), \qquad (6)$$

where $\otimes$ is a convolution, $\mathbf{k}_j$ is the position of the segment j in the $k_x$, $k_y$ plane and rect(x,y)=1 for {|x|<1/2 & |y|<1/2} and 0 elsewhere.

The excitation field $E_{in}(x,y)$ is then given by the Fourier transform of the previously expressed field

$$E_{in}(\mathbf{x}) = \sum_{j=1}^{nSeg^2} e^{i\varphi_j} \cdot \text{sinc}\left(\frac{nSeg}{NA}\frac{\mathbf{x}}{2\pi}\right)\cdot e^{-i2\pi \mathbf{x} \mathbf{k}_j} = \sum_{j=1}^{nSeg^2} e^{i\varphi_j} E_{in,j}(\mathbf{x}) . \qquad (7)$$

The overlap between the excitation field and the modes can then be written as

$$\alpha_n = \iint_{x,y} \sum_{j=1}^{nSeg^2} e^{i\varphi_j} E_{in,j}(x,y) \cdot E_n^*(x,y) dy dx = \sum_{j=1}^{nSeg^2} e^{i\varphi_j} \cdot \gamma_{n,j} \qquad (8)$$

and the field produced by the propagation can then be written as

$$E_{out}(x,y) = \sum_{n=1}^{nMode}\sum_{j=1}^{nSeg^2} e^{i\varphi_j} \cdot \gamma_{n,j} E_n(x,y) \cdot e^{i\beta_n L} = \sum_{j=1}^{nSeg^2} e^{i\varphi_j} \cdot S_j(x,y,L), \qquad (9)$$

where $S_j(x,y,L)$ is a new subspace of complex modes generated by the pupil segmentation and $\varphi_j$ the controlled phase of the $j^{th}$ segment. The output field is then described as the interference of $nSeg^2$ orthogonal fields. The wavefront (or set of $\varphi_j$) required to get constructive interference at any position ($x_0$, $y_0$) can be retrieved by following the standard wavefront-shaping procedure as described in section 3, but can also be directly expressed as $\varphi_j = -\arg[S_j(x_0, y_0, L)]$. This substitution of fiber modes by a degenerate subspace produces a degradation of the ability to focus light at any position of the distal end fiber facet, an effect similar to classical aberrations observed for standard lenses. However, unlike the aberration in lenses, the focus distortions now arise from the inability to properly address all the required modes and obey rules defined by the properties of the fiber modes in Fourier space.

Let us now briefly compare the two different approaches of light focusing through a multimode fiber described above. The first approach allows retrieving the optimal phase pattern on the fiber input after direct phase measurements, followed by the procedure of phase conjugation. Implementation of the first approach requires a more complicated experimental setup but in theoretical modeling it allows to reconstruct an ideal wavefront for focusing to a particular point on the fiber output that doesn't suffer from the segmentation at the spatial light modulator. The second approach is more practical in an experimental realization, but always limited by the number of used segments.

## 3. Experimental setup

Our experiments are performed on a conventional step-index multimode fiber (Thorlabs, FG050UGA) with a silica core of 50 μm diameter, an NA = 0.22 and a length of 15 cm. Such a fiber sustains approximately 1200 modes. We use the continuous-wave linearly polarized output of a He-Ne laser with a wavelength of 633 nm. A single-mode fiber is used to clean the laser mode and to expand the laser beam in order to match the surface of our spatial light modulator. Microscope objectives are used to couple the light into the fiber, to collect light and to image the distal end of the fiber on the camera (see Fig. 2). Before optimization of the wavefront, the light coupled into the fiber produces a speckle-like pattern on the output fiber facet. The Lee amplitude holography method [26] is used to create the desired input wavefront. To control the spatial phase of the coupled light with high speed, we use a 1920x1200 Vialux V4100 digital micromirror device (DMD). Each mirror of the DMD can be set to two different tilt angles. Together the mirrors of the DMD create a binary 2D blazed grating. Lenses L1 and L2 are placed in a 4f-configuration to image the phase mask on the back focal plane of a coupling objective. A pinhole in the Fourier plane blocks all the diffraction orders except the 1st, encoding the desired spatial phase distribution. We then divide the DMD area illuminated by the beam into a square grid of segments and individually modulate the phase of each segment by spatially shifting their grating pattern in three steps of $2\pi/3$ [0, $2\pi/3$, $4\pi/3$] each. The field from all the other segments is used as reference. For each segment, the interference between the modulated segment and all the other reference segments is measured and the phase leading to the highest intensity in the desired spot on the fiber output is kept. As a result, the intensity on the target is enhanced relative to the uncontrolled initial speckle.

Since the input beam has a round shape, the segments located in the corners of the segmentation grid cannot be coupled into the fiber. For a given total number of segments organized in a square lattice of size $nSeg \times nSeg$, the number of contributing segments is given by $\pi(nSeg/2)^2$, which corresponds to about 78% of the total number of segments.

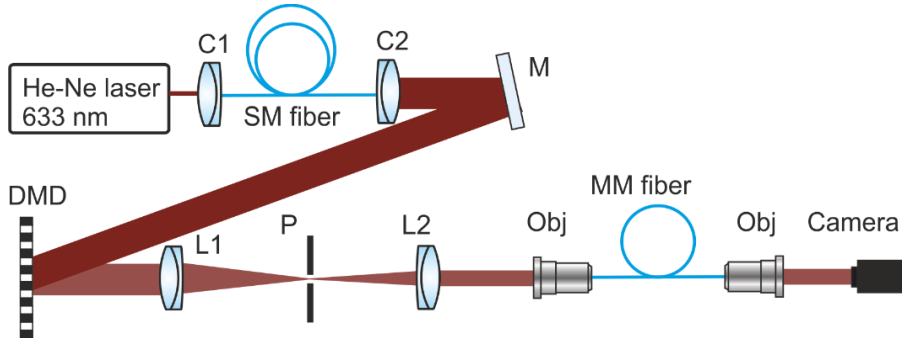

Fig. 2. Experimental setup (MM, multimode fiber; DMD, digital micromirror device; SM, single mode fiber; M, mirror; L, lenses; Obj, objectives; C, collimators; P, pinhole).

## 4. Results and discussion

Focusing light with a diffraction-limited spatial resolution at any desired position is crucial for aberration-free fluorescent imaging. To address the aberration problem in multimode-fiber imaging, we analyze the quality of the focal spots created at different distances from the fiber core axis. For focusing light through a multimode fiber, we use the standard wavefront shaping procedure described above. The optimal number of segments controlled by the spatial light modulator is chosen as a compromise between imaging contrast and optimization speed: a high number of segments leads to a high enhancement and a low speed. The wavefront-shaping procedure limits the imaging speed, because additional time is needed to extract the variety of phase masks corresponding to all desired focus positions. To characterize the imaging contrast, we define the parameter $\gamma^2 = I_{focus}/I_{total}$, where $I_{focus}$ is the power integrated over the dimensions of the focused spot and $I_{total}$ is the total power transmitted through the fiber. We experimentally show that 300 active segments give a perfect diffraction-limited spot near the fiber core axis with a $\gamma^2$ of $0.40 \pm 0.02$ (40% of the light is concentrated in the focus) that corresponds to a highly visible peak fully appropriate for imaging purposes.

In the first set of experiments, a total of 65 foci have been independently optimized at different positions on the distal end of the multimode fiber, by using 300 active segments on the DMD. To accomplish this, we simultaneously record the intensity of 65 groups of 4 by 4 binned camera pixels organized in a square lattice and extract the phases leading to the highest intensity in every of 65 positions. The spacing between two neighboring foci is about 2.5 µm. After a single optimization loop, we sequentially display every optimized phase pattern on the

DMD and image the output intensity picture with the camera. Images of all foci have been recorded with the same parameters of the camera. To demonstrate the quality of each focal spot, we display the incoherent sum of all recorded pictures in Fig. 3(a). In the center of the fiber a focus with a full width at half maximum (FWHM) of 1.43±0.06 µm is achieved, which corresponds to an effective numerical aperture of 0.22. As the focus position moves toward the fiber's cladding, the focus becomes elliptical, with the smaller axis directed toward the center of the fiber.

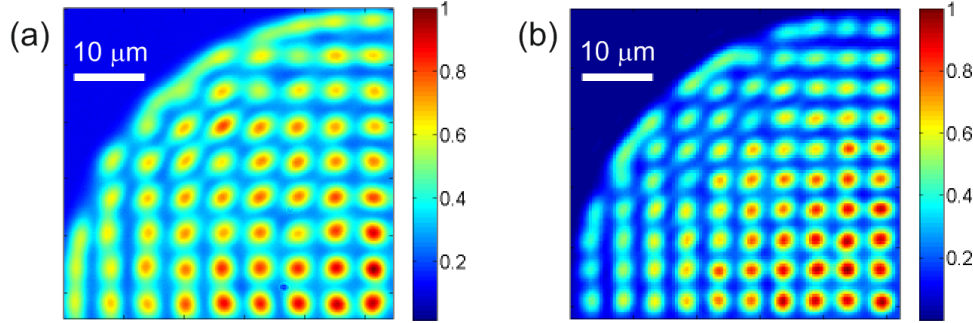

Fig. 3. Incoherent sum of the intensity images of 65 independently optimized foci at the distal end of the fiber by using 300 segments on the DMD: (a) experimental results; (b) computation using the theoretical model. The scale bars are 10µm and the incoherent sum has been normalized to the highest intensity.

The experimental results will now be compared with our theoretical model. We calculate using Eq. (9) the field on the fiber output after wavefront shaping with 300 effective DMD segments. Figure 3(b) shows the results of theoretical calculations: the incoherent sum of all independently optimized and recorded foci. The experimental results perfectly match the focus distortion predicted by our theoretical model. In the center of the fiber, diffraction-limited focusing with a FWHM of 1.42±0.04 µm and $\gamma^2$ of 0.41 is achieved. However, our calculations show that even for an ideal multimode fiber, the focus becomes elliptical near the fiber edge.

To understand the cause of this focal distortion, we analyze the structure of phase patterns required to focus light at different distances from the fiber core axis. Figure 4 shows the ideal pupil phase distributions computed by using Eq. (5) for two radial focal positions equal to 8 µm and 14.5 µm. Focusing light near the fiber core axis requires a low-spatial frequency phase pattern, as can be seen from Fig. 4(a). Setting the input field $E_{in}$ to a Dirac delta function located at the focus position near the fiber core axis, we find from the mean of Eq. (2) that in this case only about 10% of all fiber-guided modes contain 90% of the total energy transmitted through the fiber and therefore are effectively involved in the focus formation. Since only a small fraction of modes is contributing to the focus, a relatively small number of segments is enough to provide a well-focused spot. In our example, 300 active segments suffice to provide a diffraction-limited spot.

Moving the focus position towards the cladding, the effective number of modes contributing to the focus formation increases to up to 25% of the total number of guided modes, leading to a seemingly random high-spatial-frequency phase pattern (see Fig. 4(b)). To correctly represent this more complex phase mask, a higher number of segments is required. Using 300 active segments leads to an elliptical focal spot with an eccentricity of about 0.6, as can be seen from Fig. 3. Due to the significantly different number of guided modes involved in focus formation at different distances from the fiber core axis, the number of active segments on the DMD governs the level of aberrations in multimode-fiber imaging.

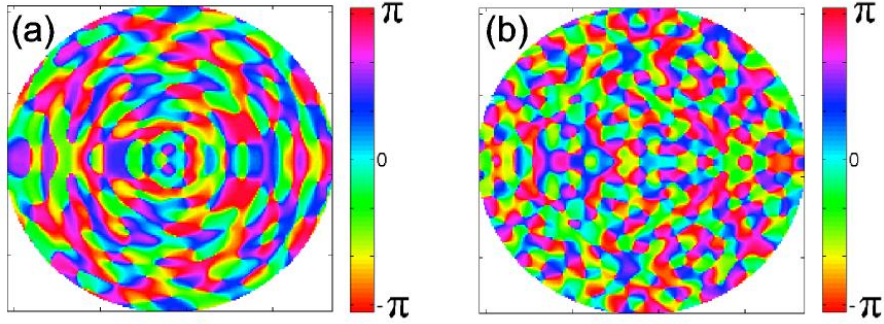

Fig. 4. Pupil phase patterns required for diffraction-limited focusing at a radial position of (a) 8 μm and (b) 14.5 μm. The radius of the phase distribution is equal to $NAk_0$. Comparing the two images we can see that making a diffraction-limited spot requires higher spatial frequencies if the target spot is further away from the center.

In the second set of experiments, we examine the quality of the focal spots created with a different number of segments on the DMD. A total of 50 foci distributed along a single radial line from the center of the fiber to the core, spaced by about 500 nm, are optimized within the same loop; therefore all the foci undergo the same level of experimental noise. After the wavefront-shaping procedure, we sequentially display every optimized phase pattern on the DMD and record the fiber output intensity with the camera. All images are then processed to extract the focus parameters. The results of the characterization of the focal spots are presented in Fig. 5, where $\gamma^2$ (Fig. 5(a)) and the FWHM in tangential direction (Fig. 5(b)) are plotted as function of the focus position along the same radial line for wavefront shaping with different number of segments. Solid lines represent the experimental data for 78 (blue curve), 300 (green curve) and 1200 (red curve) segments controlled by the DMD. The dashed lines with corresponding colors represent the result of theoretical calculations for an ideal fiber with 50μm core diameter and an NA = 0.22, averaged over 10 runs with variable fiber length (i.e. 10 different set of $S_j$, Eq. (9)).

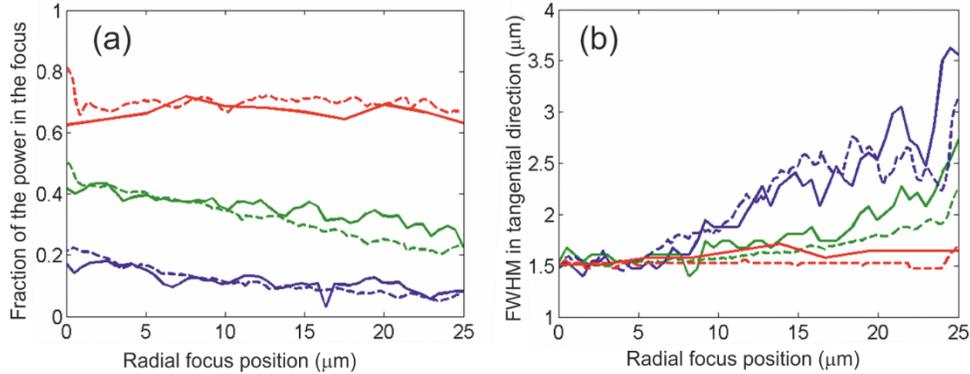

Fig. 5. Fraction of the power in the focus, $\gamma^2$ (a) and the focus FHWM in the tangential direction (b) versus the focus position along a radial line for 78 (blue curve), 300 (green curve) and 1200 (red curve) effective number of segments controlled by the DMD during wavefront shaping. Solid lines present experimental data, dashed lines the computation.

The fraction of the power in the focus grows with the number of segments used in wavefront-shaping procedure independently of the radial position, as can be seen from Fig. 5(a). For the same low number of segments, $\gamma^2$ decreases toward the edge of the fiber core. However, for 1200 segments $\gamma^2$ remains practically constant over the full area. Interestingly, the behavior of the FWHM is different. There always exists a central region of radius $r \approx 8$ μm, where perfect focusing is allowed, independent of the actual number of segments. Increasing the number of segments increases the aberration-free area, finally matching the complete core area of the fiber. Our experimental and theoretical results show that for multimode fiber with a core radius of 50 µm and NA = 0.22, aberration-free imaging is possible with 1200 or more segments.

Finally, we analyze the focus distortion inherent to multimode-fiber imaging with a low number of segments from the position of aberration theory. We consider a focus optimized using 78 segments at a distance of 18 µm from the fiber core axis. Fig 6(a) shows the intensity distribution at the distal end of the fiber after wavefront shaping, computed using our theoretical model. Strong deformation of the focus is observed including an elongation in the tangential direction with a FWHM equal to 2.9 µm, as well as an opposite and proportional contraction in the radial direction with a FWHM equal to 1.1 µm. Cross sections of the focal spot are plotted in Fig 6(b) where the black solid line corresponds to the radial direction and the blue solid line corresponds to the tangential direction. The red dashed line represents a cross section of a diffraction-limited focal spot created near the fiber core axis. Remarkably, we see in Fig 6(b) that the focus FWHM in the radial direction is smaller than the diffraction-limited focus size as the focus position moves away from the center of the fiber, leading to sub-diffraction focusing in the radial direction. The focus compression in radial direction is accompanied with the emergence of pronounced side lobes. The side lobes are clearly visible and have a central-peak-to-side-lobe ratio of 4.5.

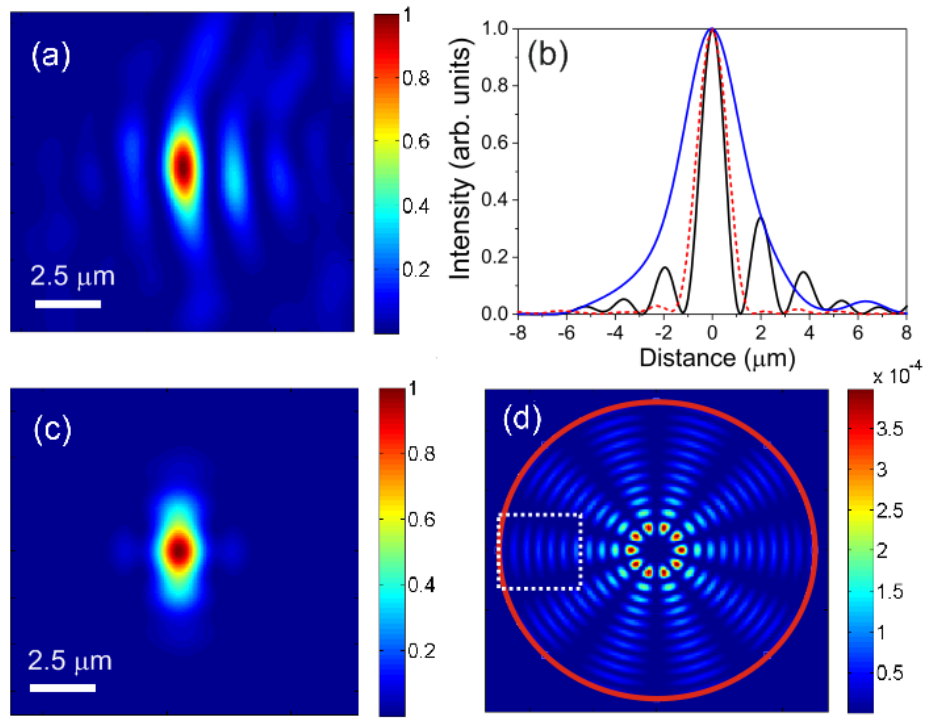

Fig. 6. (a) Intensity distribution at the distal end of the multimode fiber computed with our theoretical model for a focus optimized at a distance of 18 µm from the fiber core axis by wavefront shaping with 78 segments. (b) Cross sections of the focal spot in (a) in radial (black solid line) and tangential (blue solid line) directions. The red dashed line represents a cross section of a diffraction-limited focal spot near the fiber core axis. (c) Theoretically calculated intensity distribution of an aberration-free focus at a distance of 18 µm from the optical axis with a $Z_2^2$ Zernike mode added, showing the focal distortion typical for primary astigmatism. (d) Intensity distribution of the $LP_{5,11}$ guided fiber mode. The red circle has a diameter 50 µm and depicts the boundary of the fiber core. The white square shows the area on the output fiber facet presented in (a). Scale bars are 2.5 µm in (a) and (d).

This wavefront profile is modeled using Zernike polynomials to yield a set of fitting coefficients that individually represent different types of aberrations. Zernike coefficients are linearly independent, thus individual aberration contributions to an overall wavefront may be isolated and quantified separately. We took into account only primary aberrations [27] and found that for the focal spot presented in Fig. 6(a), a $Z_2^2$ Zernike mode has a maximal contribution to focus distortion. $Z_2^2$ Zernike mode corresponds to primary astigmatism aberration. Figure 6(c) shows the calculated intensity distribution when a $Z_2^2$ Zernike mode with maximal amplitude of $\pi$ is added to an aberration-free focus with *NA* = 0.22 at a distance of 18 µm from the optical axis. The resulting wavefront is then propagated over 2.5 µm in free space. Strong deformation of the focus is observed including an elongation in the tangential direction with FWHM equal to 2.5 µm, as well as a contraction in the radial direction with a FWHM equal to 1.4 µm and pronounced side lobes. Our analysis shows that the resulting shape of the beam suffering from astigmatism matches the shape of the focus on the fiber output well, although they arise from completely different optical systems.

Despite the similarity between the conventional astigmatism and aberrations of the focal spot near the fiber core edge after wavefront shaping, the nature of the focal distortions in these two cases is different. The shape of the aberrated focal spot in multimode-fiber imaging with a low number of segments arises from the shape of guided fiber modes. With a total of 78 segments controlled by the DMD, only about the same number of modes can be independently addressed. That makes a contribution of every mode extremely important and distorts the perfect shape of the focus accordingly to the shapes of the propagating modes. Using Eq. (2) we compute the overlap coefficients of the optimized focus created at a distance of 18 µm from the fiber core axis with all the fiber modes and retrieve the most contributing modes. Our calculations show that the $LP_{5,11}$ mode has the largest contribution to the focus from Fig 6(a). Elongation of the focal spot in the tangential direction is related to the central symmetrical field distribution of the $LP_{5,11}$ mode presented in Fig. 6(d). The presence of side lobes in the focus shape arises from the oscillating behavior of the mode. As a result, aberration in multimode-fiber imaging originates from the field distribution of the specific fiber guided mode.

## 5. Conclusions

In this work, we provide a systematic study of point-spread-function aberrations inherent to focusing through a multimode fiber. We investigated focus distortions for a variety of parameters and showed that aberration-free imaging with a large field of view requires using an appropriate number of segments on the spatial light modulator during the wavefront shaping procedure. We developed a theoretical model of wavefront shaping through a multimode fiber, which confirmed our experimental results and allowed to analyze the nature of focal distortions. Elongation of the focal spot in the tangential direction, as well as side lobes has been shown to originate from the field distribution of the specific fiber guided mode. The similarity between the conventional primary astigmatism aberration and the distortion of the focal spot near the fiber core edge has been shown. Besides providing a better understanding of the limits of wavefront shaping through a multimode fiber, these results are important for multimode-fiber imaging applications, especially there where speed is crucial.

## Acknowledgments

We thank Allard Mosk, Christoph Husemann, Joerg Petschulat, Cornelis Harteveld and Willem Vos for support and discussions. This work has been funded by FOM and NWO (Vici grant) and STW.